\documentclass[5p,numbers,sort&compress,twocolumn]{elsarticle}

\usepackage{amssymb}
\usepackage{xcolor}
\usepackage{graphicx}
\usepackage{dcolumn}
\usepackage{bm}
\usepackage{float}
\usepackage{threeparttable}
\usepackage[colorlinks = true, citecolor=blue,urlcolor=blue,linkcolor=blue]{hyperref}
\usepackage{color} 
\usepackage[T1]{fontenc}

\usepackage{url}
\usepackage{amsmath}
\usepackage{mathtools, cuted}
\usepackage[english]{babel}
\usepackage[output-decimal-marker={.},exponent-product=\cdot]{siunitx}
\usepackage[normalem]{ulem}
\journal{Physics Letters B}

\definecolor{mbscolor}{rgb}{0.60, 0.0, 0.65}

\renewcommand{\vec}[1]{{\mathbf{#1}}}
\newcommand{\vell}{\boldsymbol{\mathbf\ell}}

\begin{document}

\begin{frontmatter}

\title{Discovery of a new long-lived isomer in $^{114}$Rh via Penning-trap mass spectrometry}

\author[jyv]{M. Stryjczyk\corref{cor}}
\ead{marek.m.stryjczyk@jyu.fi}
\cortext[cor]{Corresponding authors}
\author[jyv]{A.~Jaries\corref{cor}}
\ead{arthur.a.jaries@jyu.fi}
\author[ulb]{W.~Ryssens\corref{cor}}
\ead{Wouter.Ryssens@ulb.be}
\author[lyon]{M.~Bender\corref{cor}}
\ead{bender@ipnl.in2p3.fr}
\author[jyv]{A.~Kankainen\corref{cor}}
\ead{anu.kankainen@jyu.fi}
\author[jyv]{T.~Eronen}
\author[jyv]{Z.~Ge}
\author[jyv]{I.D.~Moore}
\author[jyv]{M.~Mougeot}
\author[jyv]{A.~Raggio}
\author[jyv]{J. Ruotsalainen}

\address[jyv]{University of Jyvaskyla, Department of Physics, Accelerator Laboratory, P.O. Box 35(YFL) FI-40014 University of Jyvaskyla, Finland}
\address[ulb]{Institut d'Astronomie et d'Astrophysique, Universit\'e Libre de Bruxelles, Campus de la Plaine CP 226, 1050 Brussels, Belgium}
\address[lyon]{Universit\'e Claude Bernard Lyon 1, CNRS/IN2P3, IP2I Lyon, UMR 5822, F-69622 Villeurbanne, France}

\begin{abstract}
We report on mass measurements of three long-lived states in $^{114}$Rh performed with the JYFLTRAP Penning-trap mass spectrometer: the ground state and two isomers with estimated half-lives of about one second. The used Phase-Imaging Ion-Cyclotron-Resonance technique allowed for the discovery of a so far unknown second long-lived isomer. All three states were produced directly in proton-induced fission on a uranium target, whereas only the isomeric states were populated in the $\beta$ decay of the $^{114}$Ru ground state with spin-parity $0^+$. We propose spin-parity assignments of $(6^-)$ for the ground state, and $(3^+)$ and $(0^-)$ for the isomers. They resolve the puzzle of anomalous fission yields of this isotope despite the existing literature assigning a low angular momentum to the ground state. The experimental evidence is further supported by a detailed analysis based on mean-field calculations with the BSkG3 model. As for many other nuclei in this mass region, considering triaxial shapes is decisive for the interpretation of low-lying states of this nucleus. The discovery of a new isomer in $^{114}$Rh and our theoretical work challenge the currently adopted spin-parity assignments in this and several other odd-odd neutron-rich Rh isotopes. 
\end{abstract}

\end{frontmatter}

\section{Introduction}


Isomerism finds its origin in the suppression of electromagnetic transitions between the isomeric and lower-lying states, which can have several reasons. The most common ones are a large difference in shape or a large difference in angular momentum, either total angular momentum $J$ or its direction relative to the nuclear shape $K$, or a combination of both \cite{Walker2020,Walker22a}. Spin isomers are particularly common for deformed odd-odd nuclei that often have several very low-lying two-quasiparticle states with quite different $J$ \cite{Liu07a}, a situation different from what is found for even-even nuclei where it is often the difference in $K$ between some two-quasiparticle state and collective rotational states that leads to isomerism.
In the case of the odd-odd rhodium (${Z=45}$) isotopes, isomeric states have been observed in the majority of isotopes~\cite{NUBASE20}. In particular, on the neutron-rich side, such long-lived excited states have been confirmed as far as for $N=73, A=118$. Thus, this isotopic chain provides a way to study the phenomenon of isomerism. However, the available spectroscopic information is quite limited, with spin-parity assignments for the ground and isomeric states beyond $^{104}$Rh being mostly tentative \cite{NUBASE20,ENSDF}. Although the databases~\cite{NUBASE20,ENSDF} propose a low-spin ground state and a high-spin isomer, mostly based on simple arguments, it is generally unknown if the ground state has high or low spin.

The ground and isomeric states in neutron-rich odd-odd Rh isotopes were the subject of a recent study at the JYFLTRAP double Penning trap~\cite{Hukkanen2023}. Aside from extremely accurate values for the binding energies of these states, this measurement also resulted in a puzzling observation for $^{114}$Rh: the fission yield of its ground state far exceeded that of the isomeric state~\cite{Hukkanen2023}. Assuming the current assignments of $1^+$ and $(7^-)$ \cite{Lhersonneau2003,NUBASE20,ENSDF} for the ground and isomeric states, first identified in Ref.~\cite{Aysto1988}, this implies that fission would favor the production of low-spin states of $^{114}$Rh over the production of high-spin states. If true, this state of affairs would make this nucleus the sole known exception in the entire region as fission favors the production of high-spin states not only for all other odd-odd neutron-rich Rh isotopes \cite{Hukkanen2023}, but also for several neighboring isotopic chains such as silver, cadmium and indium \cite{Rakopoulos2019,Gao2023,Ruotsalainen2023,deGroote2023}. This anomaly would evidently disappear if the energy order of these two states of $^{114}$Rh were inverted.

Another known peculiarity of $^{114}$Rh concerns the inverted signature splitting of its negative-parity yrast band \cite{Liu11a,Navin2017} that is built on the state of $^{114}$Rh and is usually identified with its isomer \cite{NUBASE20,ENSDF}. This means that the even- and odd-spin levels in this $\Delta I = 1$ band exhibit a small relative shift into the opposite direction of what is found in most cases, and in particular for the lighter odd-odd Rh isotopes. This anomaly would disappear if the band head had an even angular momentum instead of $(7^-)$. 

We report here on Penning-trap measurements of the binding energies of multiple long-lived states in $^{114}$Rh as a follow-up experiment to Ref.~\cite{Hukkanen2023}. Our strategy hinges on the fact that different production techniques result in different yields for each long-lived state as the $\beta$ decay of the $^{114}$Ru $0^+$ ground state populates only low-spin states~\cite{Jokinen1992} while fission produces both high- and low-spin states~\cite{Lhersonneau2003,Hukkanen2023}. As demonstrated for other isotopes in the region~\cite{Hukkanen2024,Hukkanen2023,Ruotsalainen2023}, measuring the binding energies of the species produced in both ways allows to unambiguously determine whether the low-spin state is the ground state or an isomer, provided our knowledge of the decay scheme of $^{114}$Ru is complete and accurate. 

Our work uncovered the existence of an additional long-lived state in $^{114}$Rh that is populated in both production mechanisms. We report on binding energy measurements for the ground state and the two isomeric states, discuss possible spin-parity assignments and estimate half-lives of the excited states. To compliment these results, we interpret the structure of all three long-lived states with the aid of mean-field calculations based on the BSkG3 model~\cite{Grams2023}. Finally, we discuss the broad implications of our results on the structure of other odd-odd neutron-rich Rh isotopes.

\section{Experimental methods}

The mass of $^{114}$Rh was measured at the Ion Guide Isotope Separator On-Line (IGISOL) facility \cite{Moore2013,Penttila2020} in Jyv\"askyl\"a, Finland. The ions of interest were produced in 25-MeV-proton-induced fission of a 15~mg/cm$^2$-thick $^{nat}$U target. First, the fission fragments were stopped in a helium-filled gas cell operating at about 300 mbars. The thermalized ions were further transported into a sextupole ion guide \cite{Karvonen2008} and accelerated by a 30~kV potential difference to ground. Then the beam was mass-separated by a 55$^{\circ}$ dipole magnet ($m/\Delta m \approx 500$) with respect to the mass-over-charge ratio and injected into the gas-filled radio-frequency quadrupole cooler-buncher~\cite{Nieminen2001}. From there, the bunched beam was delivered to the JYFLTRAP double Penning-trap mass spectrometer \cite{Eronen2012}. 

In the first trap, the mass-selective buffer-gas cooling technique \cite{Savard1991} was applied. There, all ions were first excited to a larger radius via dipolar magnetron excitation, and only the ions of interest were re-centered to be further extracted to the second trap for the mass measurement.
This method was used for the $^{114}$Rh$^+$ states populated directly in fission, measured with $^{85}$Rb$^{+}$ as a reference ion. To study the states produced via the $\beta$ decay of $^{114}$Ru, two different schemes were used for the ions' preparation. In the first scheme, which was used for the measurements against the singly-charged $^{85}$Rb$^{+}$ reference, the $^{114}$Ru$^{+}$ ions were trapped and cooled in the first trap for about 160 ms to let them decay (${T_{1/2} = 540(30)}$~ms \cite{NUBASE20}). The buffer-gas cooling technique was then applied to the produced doubly-charged $^{114}$Rh$^{2+}$ daughter ions. Once centered, they were transferred to the second (measurement) trap for about one period of magnetron motion (600~$\mu$s). After this, the $^{114}$Rh$^{2+}$ ions were transferred back to the first trap for an additional centering and cooling. The second scheme was used for the measurements of the two populated $^{114}$Rh$^{2+}$ states against each other. It was otherwise similar to the first one except an extended waiting time to 360 ms to allow for more decays of $^{114}$Ru. 

\begin{figure}[h!t!b]
    \centering
    \includegraphics[width=0.9\columnwidth]{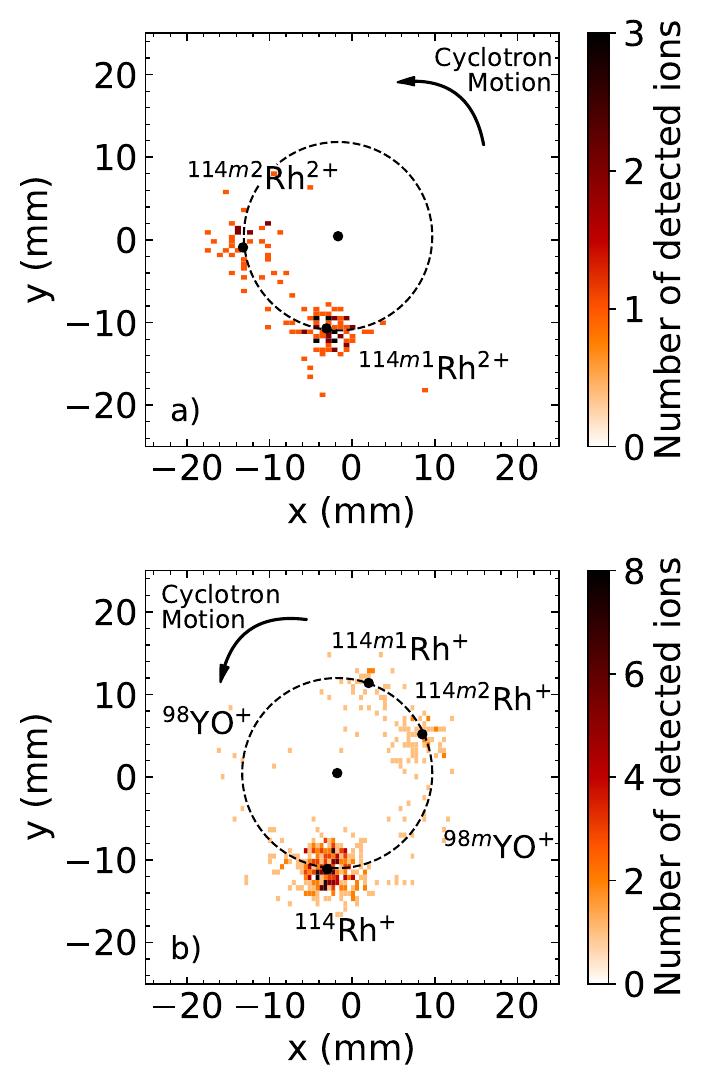}
    \caption{\label{fig:piicr}a) Projection of the cyclotron motion of the two long-lived excited states in $^{114}$Rh produced via in-trap decay onto the position-sensitive detector obtained with the PI-ICR technique using a phase accumulation time $t_{acc} = 550$~ms. b) Projection of the cyclotron motion of the three states in $^{114}$Rh and isobaric contaminants produced in fission onto the position-sensitive detector obtained with the PI-ICR technique using a phase accumulation time $t_{acc} = 644$~ms. A gate on the count rate is applied on both panels, up to 1 detected ion per bunch.}
\end{figure}

After transfer to the second trap, the mass-over-charge ratio of the ions of interest was determined using the Phase-Imaging Ion-Cyclotron-Resonance (PI-ICR) technique \cite{Eliseev2013,Eliseev2014,Nesterenko2018,Nesterenko2021} by measuring the ions' cyclotron frequency ${\nu_c = \frac{1}{2\pi}\frac{B}{(m/q)}}$ in a magnetic field $B$. This frequency is obtained from the phase differences between the ion's cyclotron and magnetron in-trap motions acquired during a phase accumulation time $t_{acc}$. Such a time was selected that the cyclotron motion projections of the ground and isomeric states and possible contaminant ions did not overlap, see Fig.~\ref{fig:piicr}.

The magnetic field strength $B$ is determined precisely by measuring a cyclotron frequency of reference ions $\nu_{c,\mathrm{ref}}$. For the majority of the cases reported in this work, singly-charged $^{85}$Rb (${\mathrm{ME}_{lit.} = -82167.341(5)}$~keV \cite{AME20}) delivered from the IGISOL offline surface ion source \cite{Vilen2020} was used as a reference. Only for the $^{114}$Rh$^{2+}$ states produced in the in-trap decay of $^{114}$Ru$^{+}$, the m1 state was used as a reference for the measurement of the m2 state, see Fig.~\ref{fig:piicr}. To account for the fluctuations of the magnetic field in time, the measurements of the ion of interest and the reference ion were alternated. The atomic mass $M$ is connected to the frequency ratio $r=\nu_{c,\mathrm{ref}}/\nu_{c}$ between the reference ions and the ions of interest:
\begin{equation} \label{eq:mass}
M = \frac{z_\mathrm{ioi}}{z_\mathrm{ref}} (M_\mathrm{ref} - z_\mathrm{ref} m_{e}) r + z_\mathrm{ioi} m_{e} + \Delta B_e\mathrm{,}
\end{equation}
where $m_e$ and $M_\mathrm{ref}$ are the mass of a free electron and the atomic mass of the reference, respectively, $z_\mathrm{ioi}$ and $z_\mathrm{ref}$ are charge states of the ion of interest and the reference ion, respectively while $\Delta B_e$ represents the atomic electron binding energy difference. We note that even for the $2^+$ ions the $\Delta B_e$ value is of the order of a few eV and, thus, it was neglected. For the in-trap-decay-produced ions, the reference ion and the ion of interest are isobaric species with the same charge state $z$. The energy difference between them, $\Delta E$, was extracted as follows:
\begin{equation}
\label{eq:Q}
\Delta E = (r-1)[M_\mathrm{ref} - zm_e]c^2 \mathrm{,}
\end{equation} 
with $c$ being the speed of light in vacuum. 

A count-rate class analysis \cite{Kellerbauer2003,Roux2013,Nesterenko2021} was statistically feasible for the ground-state measurement with $t_{acc}=644$~ms. For all the other cases, the count rate was limited to one detected ion per bunch. During the analysis, the systematic uncertainties related to the magnetron phase advancement, the angle error and the temporal magnetic field fluctuation were taken into account \cite{Nesterenko2021}. In addition, for the cases measured against the $^{85}$Rb$^+$ reference ions, a mass-dependent uncertainty of $\delta r/r = -2.35(81) \times 10^{-10} / \textnormal{u} \times (M_\mathrm{ref} - M)$ and a residual systematic uncertainty of $\delta r/r=9\times 10^{-9}$ were added \cite{Nesterenko2021}. 

\section{\label{Sec:results}Results}

\begin{table*}
\centering\scriptsize
\begin{threeparttable}
\caption{\label{tab:results} The frequency ratios ($r=\nu_{c, \rm ref}/\nu_{c}$), corresponding mass-excess values (ME) and isomer excitation energies ($E_{x}$) measured in this work using the listed reference ions (Ref.). The charge state of the ion of interest $z_\mathrm{ioi}$ and the reference $z_\mathrm{ref}$ as well as the used accumulation times $t_{acc}$ are also listed. The literature mass-excess values (ME$_{lit.}$) and isomer excitation energies ($E_{x, lit.}$) are taken from Ref. \cite{Hukkanen2023}. The differences ${\mathrm{Diff.} = \mathrm{ME}-\mathrm{ME}_{lit.}}$ are provided for comparison. The state order, tentative spin-parity assignments $J^{\pi}$ and half-lives $T_{1/2}$ are deduced in this work, see text for details.}
\begin{tabular}{lllllllllllll}
\hline
Nuclide & $J^{\pi}$ & $T_{1/2}$ & Ref. & $z_\mathrm{ref}$ & $z_\mathrm{ioi}$ & $t_{acc}$ & $r=\nu_{c, \rm ref}/\nu_{c}$ & ME & ME$_{lit.}$ & $E_x$ & $E_{x, lit.}$ & Diff.\\ 
 & & (s) &  &  & & (ms) & & (keV) & (keV) &  (keV)  & (keV) & (keV) \\\hline
$^{114}$Rh          & $(6^-)$   & 1.85(6)\tnote{a}     & $^{85}$Rb & 1    & 1 & 644 & \num{1.341615348(24)}  & \num{-75662.5(19)} & \num{-75662.7(26)}   & &  & $0.2(32)$ \\[0.15cm]
$^{114}$Rh$^{m1}$   & $(0^-)$   & $\sim 1$     & $^{85}$Rb & 1     & 1 & 644 & \num{1.341616521(23)}  & \num{-75569.7(18)} & & & & \\
                    &           &               & $^{85}$Rb & 1     & 2 & 550 & \num{0.670805047(9)}  & \num{-75567.0(14)} &  & &  &  \\
 \multicolumn{8}{r}{Final result:}  & \num{-75568.0(13)} & \num{-75551.8(41)}\tnote{b} & $94.5(23)$ & $110.9(32)$\tnote{b} & $-16(4)$ \\[0.15cm]
$^{114}$Rh$^{m2}$   & $(3^+)$     & $\sim 1$     & $^{85}$Rb  & 1    & 1 & 644 & \num{1.341616817(21)}   & \num{-75546.3(17)} &  & &  &  \\
   &      &     & $^{85}$Rb & 1 & 2 & 550 & \num{0.670805191(9)}  & \num{-75544.3(15)} &  & &  &  \\
   &      &     & $^{114}$Rh$^{m1}$ & 2 & 2 & 480 & \num{1.000000218(9)}\tnote{c}  & \num{-75545.6(16)} &  & &  &  \\
 \multicolumn{8}{r}{Final result:}  & \num{-75545.3(9)} & \num{-75551.8(41)}\tnote{b}  & $117.2(21)$ & $110.9(32)$\tnote{b} & $6.5(42)$ \\\hline
\end{tabular}
\begin{tablenotes}
\item[a]{Proposed in this work based on Ref. \cite{Lhersonneau2003}}
\item[b]{Reported in Ref. \cite{Hukkanen2023} as $^{114}$Rh$^{m}$.}
\item[c]{This frequency ratio corresponds to the energy difference of $22.4(9)$ keV between $^{114}$Rh$^{m2}$ and $^{114}$Rh$^{m1}$, as calculated using Eq.~\ref{eq:Q}.}
\end{tablenotes}
\end{threeparttable}
\end{table*}

Two states were observed in the in-trap decay of the $^{114}$Ru($0^+$) ground state, see Fig. \ref{fig:piicr}a). Five different accumulation times were tested: 220, 420, 480, 550 and 620~ms, with 550~ms being chosen for the final measurement against $^{85}$Rb. An additional measurement to extract the energy difference between these two states was performed with ${t_{acc} = 480}$~ms using the more bound state as a reference. The extracted mass-excess values, $-75567.0(14)$~keV and $-75544.3(15)$~keV, do not agree with neither $-75662.7(26)$~keV nor $-75551.8(41)$~keV reported in Ref. \cite{Hukkanen2023} for the $^{114}$Rh ground state and isomer, respectively. The ratios of the numbers of ions of the less to the more bound state for different measurement cycles are similar and are about $2:1$. 

In the case of the $1^+$ ions produced directly in fission, the measurement was performed against $^{85}$Rb$^{+}$ with ${t_{acc} = 644}$~ms. The three observed states (see Fig. \ref{fig:piicr}b) have mass-excess values of $-75662.5(19)$~keV, $-75569.7(18)$~keV and $-75546.3(17)$~keV, respectively. The mass of the most bound state agrees with the ground-state mass of $^{114}$Rh reported in Ref. \cite{Hukkanen2023} (${\mathrm{ME}_{lit.} = -75662.7(26)}$~keV) while the mass values of the two other states agree with the values measured using the $2^+$ ions.
The summary of the measured frequency ratios, deduced mass-excess values and a comparison with Ref. \cite{Hukkanen2023} is presented in Table \ref{tab:results}. Based on these reported measurements, we concluded there are three long-lived states present in $^{114}$Rh, the ground state and two isomeric states at 94.5(23)~keV and 117.2(21)~keV excitation energies, averaged on the different measurements.  

\begin{figure}[h!t!b]
    \centering
    \includegraphics[width=\columnwidth]{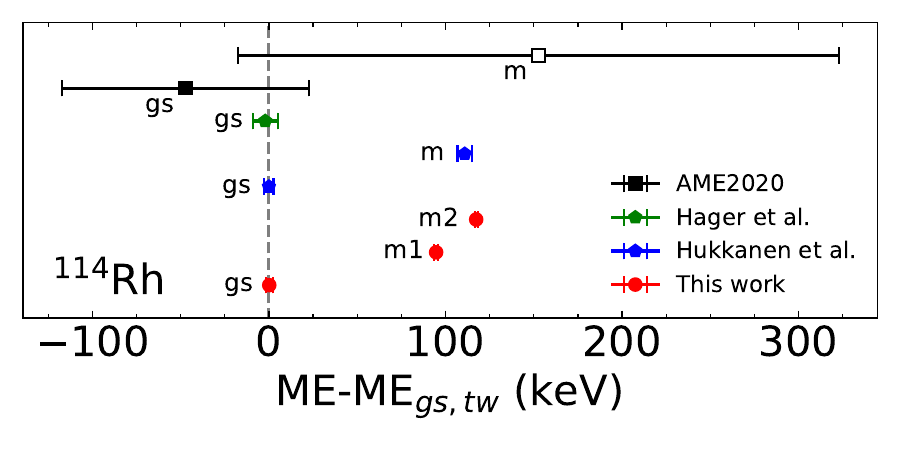}
    \caption{\label{fig:114h_states_comp} Comparison between the mass-excess values in keV of the three long-lived states in $^{114}$Rh measured in the work (in red) and the previous JYFLTRAP results, Hukkanen \textit{et al.} \cite{Hukkanen2023} in blue and Hager \textit{et al.} \cite{Hager2007} in green, with respect to the mass excess of the ground state from this work (ME$_\text{gs,tw}$). The evaluated mass-excess values from AME20 \cite{AME20} and NUBASE20 \cite{NUBASE20} are also shown in black. An open black marker is used for the mass of the isomeric state based on extrapolations.}
\end{figure}

The most likely explanation why the two isomers were not observed in Ref. \cite{Hukkanen2023} is a too short accumulation time applied during the measurement. This hypothesis is supported by the fact the mass excess reported in Ref.~\cite{Hukkanen2023} lies between the mass-excess values of the two isomers reported in this work, see Fig. \ref{fig:114h_states_comp}.

We note that in the fission data there are two other small ion clusters present. They are most likely the isobaric contamination of $^{98gs,m}$Y$^{16}$O. The YO$^+$ molecules are known contaminants at IGISOL and $^{98}$Y is one of the strongest produced isotopes of yttrium in fission \cite{Penttila2016}. Since these species are only 12~Hz from the ion of interest, they are not fully separated in the first trap. The mass-excess of the stronger produced $^{98m}$Y$^{16}$O is $-76565.1(19)$~keV compared to $\mathrm{ME}_{lit.} = -76560(8)$~keV \cite{NUBASE20}.

The ground state of $^{114}$Rh is produced only in fission and it is the strongest-produced state, with the ratio between the number of registered ions in $^{114}$Rh$^{gs}:^{114}$Rh$^{m1}:{}^{114}$Rh$^{m2}$ being ${15:1:2.6}$. At the same time, the $^{114}$Rh$^{m1}:{}^{114}$Rh$^{m2}$ ratio in the in-trap decay is, as mentioned earlier, about ${2:1}$. Considering that the production of higher-spin states is favored in fission while population of lower-spin states can be expected from the decay of $^{114}$Ru($0^+$), we can conclude that the ground state is the high-spin state ($^{114}$Rh$^\text{hs}$), the first isomeric state is the low-spin state ($^{114}$Rh$^\text{ls}$) while the second isomeric state is the medium-spin state ($^{114}$Rh$^\text{ms}$). This order is reversed with respect to the literature \cite{NUBASE20} where the high-spin state was proposed to be isomeric.

\subsection{Half-lives}

The half-lives of the long-lived states in $^{114}$Rh are known mostly from Ref. \cite{Lhersonneau2003}. However, considering that the beam contained the entire $A=114$ isobar, these values might not be reliable. The weighted average of the partial half-lives determined from the $\gamma$ rays deexciting the $J \geq 6$ states at 1501, 1984, 2520, 2598 and 2623 keV, as reported in Ref. \cite{Lhersonneau2003}, is 1.85(6)~s. Since these are high-spin states, it is unlikely they are fed directly or indirectly in the $\beta$ decay of $^{114}$Rh$^\text{ls,ms}$. In addition, $^{114}$Rh$^\text{hs}$ is produced only directly in fission, thus, the $^{114}$Ru contamination in the beam does not influence the value, as it was observed in the case of the half-life of the low-spin state in $^{112}$Rh \cite{Hukkanen2023}. Consequently, we adopt $1.85(6)$~s as the half-life value of $^{114}$Rh$^\text{hs}$. 

While an unambiguous determination of the $^{114}$Rh$^\text{ls,ms}$ half-lives based on the values reported in Ref. \cite{Lhersonneau2003} is hindered due to possible direct and indirect feedings from the decays of $^{114}$Rh$^\text{hs}$ and $^{114}$Ru, an estimation can be made. The partial half-life deduced from the $\beta$-delayed 783-keV $\gamma$-ray transition de-exciting the ($0^+$) state at 1116 keV in $^{114}$Pd is 1.65(35)~s \cite{Lhersonneau2003}. Due to a relatively high energy and a low spin value, this state is, most likely, almost exclusively populated in the $\beta$ decay of $^{114}$Rh$^\text{ls}$. However, the value might be overestimated as $^{114}$Rh$^\text{ls}$ is produced in the $\beta$ decay of $^{114}$Ru. A similar observation was made for the determination of the half-life of the low-spin state in $^{112}$Rh \cite{Hukkanen2023}. From the mass measurements reported in this work, the low- and the medium-spin states are known to have half-lives long enough to survive the 1.18~s fission cycle and 0.9~s in-trap decay cycle. In addition, the ion ratio between the isomers was rather constant in time for the used accumulation times between 220 and 620 ms, indicating that the decay constants are of the same order of magnitude. Based on this information, we can estimate the half-lives of $^{114}$Rh$^\text{ls,ms}$ to be of the order of one second. 

The low-spin isomeric states of $^{104-112}$Rh are typically shorter-lived than the high-spin states \cite{NUBASE20}. Shorter half-lives of the low- and medium-spin isomeric states might also explain why the mass-excess value reported in Ref.~\cite{Hager2007}, measured with 800~ms excitation time, agrees with the ground-state mass value of $^{114}$Rh from Ref. \cite{Hukkanen2023} as well as this work, see Fig. \ref{fig:114h_states_comp}. In addition, they provide constraints for the spin assignments. Based on our estimates it can be concluded that the spin difference between each pair of the long-lived states should be at least 3$\hbar$ as a lower value would enable an $M2$ internal transition resulting in a much faster decay. 

\subsection{Spin-parity assignments}

The two known long-lived states in $^{114}$Rh were assigned as $1^+$ for the low-spin state \cite{Aysto1988,Jokinen1992} and $(7^-)$ for the high-spin state \cite{Lhersonneau2003,Liu11a,Navin2017}. In Refs. \cite{Liu11a,Navin2017} the assignment for the high-spin state is tentative and based on similarities with adjacent Rh isotopes \cite{Liu11a} as well as theoretical calculations \cite{Navin2017}, while in the $^{114}$Rh decay study \cite{Lhersonneau2003} it is based on the decay pattern. However, we note that there was no direct $\beta$ feeding observed to any known $8^-$ or $9^-$ states. At the same time, a strong $\beta$ feeding to the $6^-$ state at 2623 keV and smaller feedings to other $6^-$ and $7^-$ states were reported \cite{Lhersonneau2003}. This points to the $6^-$ spin-parity assignment for the $^{114}$Rh$^\text{hs}$ ground state. A small feeding of the $8^+_1$ state in $^{114}$Pd might be explained by the presence of a pandemonium effect \cite{Hardy1977a}. 

Considering that the spin difference between the states should be at least three, the low- and medium-spin states can be assigned spin 0 and 3, respectively. To assign the parity for the low- and medium-spin states, we used the $2:1$ ratio of the $^{114}$Rh$^\text{ls}:{}^{114}$Rh$^\text{ms}$ ions produced in the in-trap decay of $^{114}$Ru. For the spin-0 level, the positive parity is unlikely as it would allow for a strong direct feeding via an allowed $\beta$ decay from $^{114}$Ru($0^+_\text{gs}$), resulting in a much higher $^{114}$Rh$^\text{ls}:{}^{114}$Rh$^\text{ms}$ ratio. At the same time, with the $0^-$ assignment the direct population is hindered as it requires a first-forbidden transition. We note that the observed $^{114}$Rh$^\text{ls}:{}^{114}$Rh$^\text{ms}$ ratio also supports the rejection of the $1^+$ assignment for the low-spin state as it would allow for a strong direct population of this state via an allowed $\beta$ decay. The medium-spin state can be populated only indirectly from higher-lying states. In this case, assigning positive parity is favored as it allows for a direct deexcitation of the strongly $\beta$-fed $1^+$ states via a single $E2$ $\gamma$-ray. These $\gamma$ rays can compete with the $E1$ transitions to the $0^-$ state. If it had negative parity, the medium-spin state could only be populated via cascades of at least two $\gamma$ rays which would significantly decrease its relative production rate.

\section{Discussion}

In the absence of known shape coexistence phenomena in this mass region, the low-lying long-lived states can be expected to be spin isomers, a common feature of many deformed odd-odd nuclei \cite{Liu07a}. The structure of such states can be analyzed and interpreted within a mean-field model. For an odd-odd nucleus like $^{114}$Rh, both the ground state and the isomer take the form of a two-quasiparticle (2qp) state, with one unpaired Bogoliubov quasiparticle for protons and neutrons each \cite{Walker22a,NR95a,Liu07a}. While the angular momenta of all paired nucleons pairwise couple to zero, those of the unpaired proton and neutron in general do not and thereby determine the total angular momentum of the nuclear configuration.

For our analysis we employ BSkG3, a recent model based on mean-field calculations with an energy density functional (EDF) of the Skyrme type~\cite{Grams2023}. It is the latest entry in the Brussels-Skyrme-on-a-Grid series, a set of large-scale models aiming to provide nuclear data for astrophysical simulations. The fitting protocol of these models included essentially all known nuclear masses, such that BSkG1, BSkG2 and BSkG3 can claim excellent root-mean-square deviations (below 800 keV) on nuclear binding energies \cite{scamps2021,ryssens2022a}. Compared to the earlier BSk models~\cite{goriely2016} and other large-scale models in the literature, the BSkG-series exploit more fully the concept of spontaneous symmetry breaking. These models are unique in their reliance on a complete three-dimensional numerical representation of the atomic nucleus, a representation which naturally allows a nucleus to exhibit triaxial deformation. This freedom is of particular interest to $^{114}$Rh and neighboring nuclei as this is a region ($Z \sim 44$, $60 \lesssim N \lesssim 78$) where both experiment~\cite{doherty2017,hakala2011,srebrny2006,svensson1995} and several models~\cite{scamps2021,goriely2009,abusara2017,moller2006,zhang2015,bucher2018} indicate that nuclear ground states break axial symmetry. We already employed BSkG1 and BSkG2 to interpret new experimental data for neutron-rich Ru and Rh isotopes in Refs.~\cite{Hukkanen2023,Hukkanen2023a} and we opt here for BSkG3 as being the most refined model. Nevertheless, these refinements mostly concern properties of finite nuclei and dense matter that are not relevant to this study. We have checked that employing the older models would not have changed any part of our discussion, although details, such as the exact deformation of the ground state of $^{114}$Rh, can differ.

\begin{figure}
    \centering
    \includegraphics[width=\columnwidth]{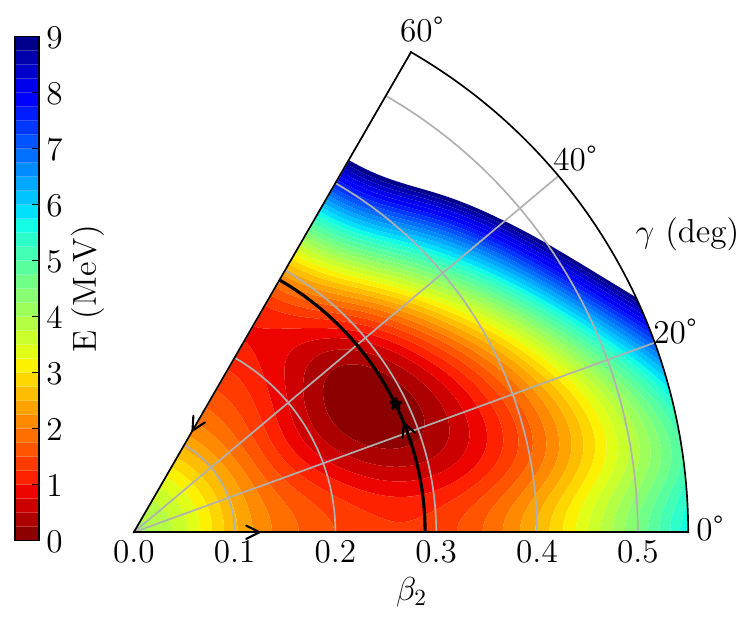}
    \caption{\label{fig:PES}Potential energy surface in the $(\beta,\gamma)$ plane for FV calculations (see text) of $^{114}$Rh with BSkG3. The trajectory followed by the Nilsson diagram in Fig.~\ref{fig:Nilsson} is indicated by black arrows. The location of the minimum obtained in a complete blocked calculation of $^{114}$Rh is indicated by a black star.}
\end{figure}

Figure~\ref{fig:PES} shows the potential energy surface (PES) of $^{114}${Rh} as a function of the total quadrupole deformation $\beta$ and the triaxiality angle\footnote{We remind the reader that the configurations with $\gamma = 0^{\circ}$ or $60^{\circ}$ correspond to prolate and oblate shapes, respectively. All other values of $\gamma$ indicate finite triaxial deformation. For the precise definition of $(\beta, \gamma)$ we refer the reader to Ref.~\cite{Hukkanen2023}.} $\gamma$ as obtained with so-called false-vacuum (FV) calculations with BSkG3. In FV calculations, one treats the nucleus as if it were even-even although the particle numbers are constrained to take their physical (odd) values $Z = 45$ and $N  = 69$. This way we can study the overall behavior of low-lying configurations while avoiding the complicated construction of 2qp configurations that correspond to the physical states for odd-odd nuclei. Qualitatively, the PES is very similar to those obtained with BSkG1 for $^{112}${Rh}~\cite{Hukkanen2023} and for $^{115}${Ru} with BSkG2~\cite{Hukkanen2023a}. Each of these calculations exhibits a minimum at a slightly different combination of $\beta$ and $\gamma$, but this is mainly due to the difference in particle numbers.

\begin{figure*}
    \centering
    \includegraphics[width=\textwidth]{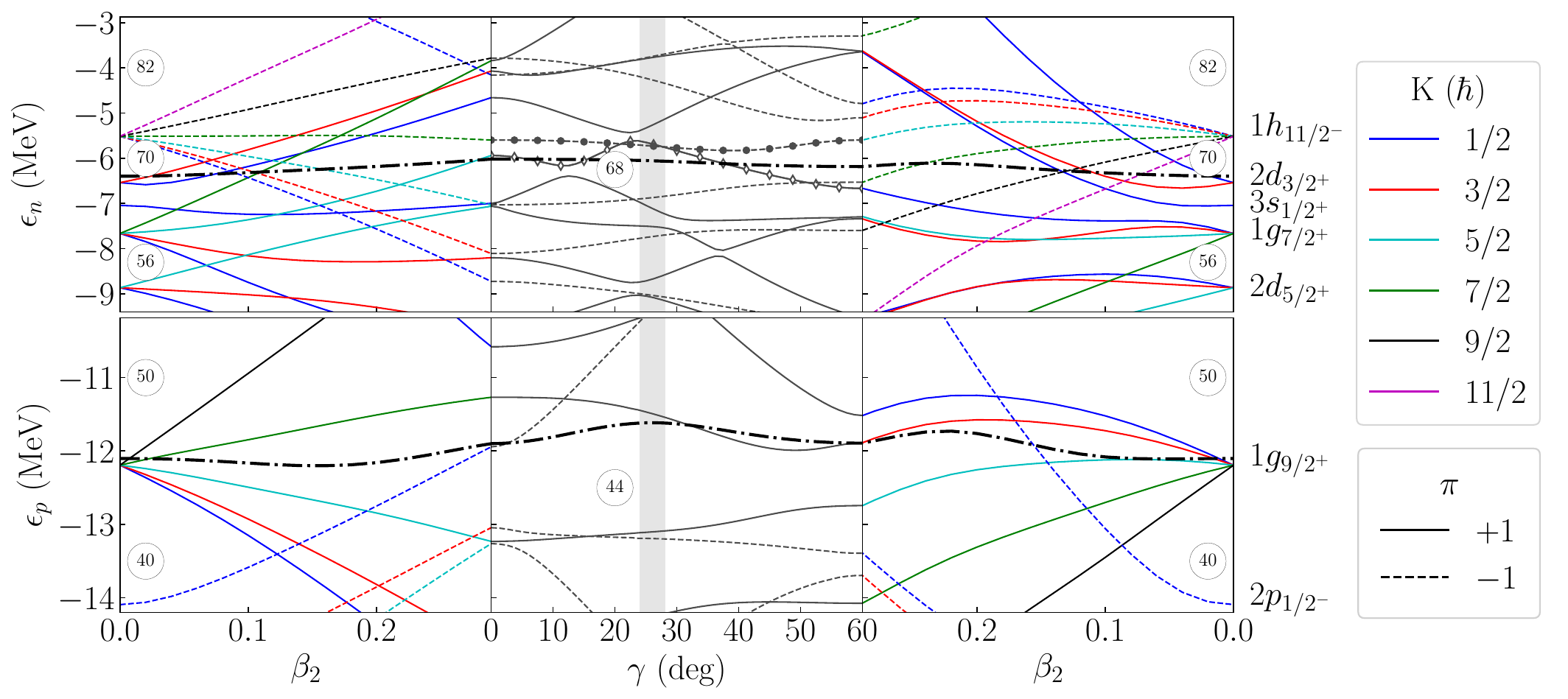}
    \caption{\label{fig:Nilsson}Eigenvalues of the single-particle Hamiltonian for neutrons (top row) and protons (bottom row) in $^{114}$Rh obtained with BSkG3 along the path in the $(\beta-\gamma)$ plane indicated by arrows on Fig.~\ref{fig:PES}. The Fermi energy is drawn as a dash-dotted line, whereas full (dashed) lines indicate single-particle levels of positive (negative) parity. The three indicated regions correspond to axially symmetric prolate shape with $\gamma = 0^{\circ}$ (left column), fixed total quadrupole deformation $\beta_{2} = 0.289$ with varying $\gamma$ (center column), and axially oblate shape with $\gamma = 60^{\circ}$ (right column). The vertical gray band in the center panels is centered at $\gamma = 26^{\circ}$, the value obtained in a complete, i.e. blocked, calculation of $^{114}$Rh. The quantum numbers of the shells at sphericity are indicated on the right-hand side. Two neutron levels near the Fermi energy are highlighted by markers in the middle column: these are the positive and negative parity levels referred to in the text as $|\diamond \rangle$ and $| \bullet \rangle$, respectively.}
\end{figure*}

Figure~\ref{fig:Nilsson} displays the Nilsson diagram of eigenvalues of the single-particle Hamiltonian of $^{114}${Rh} calculated with BSkG3 along a closed path of FV configurations in the $\beta$-$\gamma$ plane as indicated in Fig.~\ref{fig:PES}. Overall, it is similar to the Nilsson diagrams of $^{112}${Rh}~\cite{Hukkanen2023} and $^{115}${Ru}~\cite{Hukkanen2023a} obtained with BSkG1 and BSkG2, respectively. There are differences in detail such as the precise placement of neutron levels when $\gamma \in ]0^{\circ}, 60^{\circ}[$. These mainly originate in the different value of $\beta$ at which the spectrum is drawn for each of these nuclei. One chief observation of Refs.~\cite{Hukkanen2023,Hukkanen2023a} is also valid here: the sparsity of the density of proton levels around the Fermi energy at $\gamma \simeq 30^{\circ}$ is what drives many $Z \simeq 44$ nuclei towards triaxial shapes. For neutrons, the change in level density around the Fermi energy when going from axial to triaxial shapes is less pronounced but still visible.

To construct low-lying 2qp configurations, we scan the Nilsson diagram for proton and neutron single-particle states near their respective Fermi energies at the deformation of the minimum of the FV PES. 

Before addressing the complexity of attributing quantum numbers to the 2qp states of nonaxial nuclei, we recall the phenomenology of 2qp states as described by a mean-field model for the more straightforward case of a well-deformed axial odd-odd nucleus. In such a case symmetry implies that the total angular momentum $\vec{j}_{p/n}$ of each of the two unpaired nucleons is aligned with the symmetry axis that we fix along the $z$ axis, $\vec{j}_{p/n} = \pm {j}_{p/n} \vec{e}_z$. Depending on whether the angular momenta of the unpaired nucleons are parallel or anti-parallel, one can produce two distinct nuclear configurations with an angular momentum $J_{2qp}$ that is either ${j}_p + {j}_n$ or $|j_p - j_n|$ and that are still aligned with the symmetry axis, such that the $z$ component of angular momentum is given by $K_{2qp} = J_{2qp}$. Because of the spin-dependence of the proton-neutron interaction, these two states slightly differ in energy~\cite{Gallagher58a}. Typically, the favored configuration is more bound by 100 keV. The angular momenta $J_{2qp}$ of these two configurations are in general quite different, one being often much larger than the other. The large difference in angular momentum between these two states (and with other levels possibly lying in between the two) can hinder electromagnetic transitions, thereby creating a long-lived spin isomer \cite{Walker22a,Liu07a}.

According to the empirical Gallagher-Moszkowski (GM) rules~\cite{Gallagher58a}, the lower of the two partnered 2qp states has aligned proton ($\vec{s}_p$) and neutron ($\vec{s}_n$) spins ($\vec{s}_p \cdot \vec{s}_n > 0$), whereas the higher 2qp configuration usually has anti-aligned spins ($\vec{s}_p \cdot \vec{s}_n < 0$). Depending on how spin $\vec{s}_{p/n}$ and orbital angular momentum $\vell_{p/n}$ are coupled to the total angular momentum $\vec{j}_{p/n} = \vell_{p/n} + \vec{s}_{p/n}$ of each of the two nucleons, the one energetically favored can be either the 2qp configuration with low $J_{2qp}$ or the one with high $J_{2qp}$.
 
The idealized textbook case sketched above, for which the angular momenta of the 2qp configurations are completely determined by $\vec{j}_p$ and $\vec{j}_n$, applies only to axially symmetric configurations for which $\vec{j}_p$ and $\vec{j}_n$ are aligned with the symmetry axis. For non-axial configurations such as the ones studied here, the angular momenta of the blocked quasiparticles are in general neither parallel, nor aligned with any given axis. In such case, the mean-field expectation values of the component of the angular momentum operator on the quantization ($z$) axis will not correspond to the actual length of the angular momentum vector and will not take a half-integer value. The expectation values of all angular momentum operators reflect that the mean-field configuration is actually a superposition of states with different angular momenta. The construction of states with good angular momentum would either require explicit projection, which is out of the scope of the BSkG3 model, more phenomenological approaches like a particle-core-coupling model \cite{Rohozinski11a,Joshi2004,Vaman2004} or the so-called projected shell model \cite{Palit17a}. The latter models, however, have parameters that are to be adjusted locally to data and therefore are of limited predictive power for the purpose of our paper. Within the mean-field model we use to analyze the data, $J_{2qp}$ can only be estimated from the expectation values for $\langle \hat{\jmath}_p \rangle$, $\langle \hat{s}_p \rangle$, $\langle \hat{\jmath}_n\rangle$, and $\langle \hat{s}_n \rangle$ as obtained from the EDF calculation with some uncertainty. 

For $^{114}$Rh, the PES of false vacua is sufficiently stiff around its minimum so that it can be expected all low-lying 2qp states have a very similar deformation. The identification of the relevant proton level is straightforward, only the one connecting to the $K = 7/2^+$ substate of the $1g_{9/2}$ on the prolate side is near the Fermi energy. At the deformation of the minimum of the PES, this state's expectation value of $\langle \hat{\jmath}_{p,z} \rangle \simeq 3.29$ indicates that it still is dominated by a $K = 7/2^+$ component. Its spin expectation value of $\langle \hat{s}_{p,z} \rangle \simeq 0.5$ indicates aligned spin and orbital angular momentum, as expected for a level from the $1g_{9/2}$ shell. This prediction of the BSkG3 model is consistent with the experimental ground-state spin of $J = 7/2^+$ of the adjacent odd-mass Rh isotopes. 

Limiting ourselves to possible candidates for low-lying 2qp states, there are only two neutron levels sufficiently close to the Fermi energy. These are marked with dots and diamonds in the Nilsson diagram. The one labeled with dots is closest and connected to the $K = 7/2^-$ level emanating from the $1h_{11/2}$ shell on the prolate side. With $\langle \bullet | \hat{\jmath}_{n,z} | \bullet \rangle \simeq 3.3$ near the minimum of the PES, this state seems not to be strongly  affected by triaxiality either. Spin and orbital angular momentum of this state are again largely aligned. Coupling it to the proton level, one obtains either $7^-$ or $6^-$ for the favored 2qp state and $0^-$ for the excited partner state, which fits the proposed quantum numbers of the ground state and the excited level at 94.5 keV. This is also the overall lowest blocked configuration found with BSkG3. Its deformation is indicated by the black star in Fig.~\ref{fig:PES}.

Another pair of 2qp states can be constructed by assigning the odd neutron to the level of positive parity that is marked with diamonds in the Nilsson diagram. At the deformation of the PES's minimum, this level is almost degenerate with the one discussed above. It therefore can be expected that the two partner states built from this neutron level will differ by (at most) a few 100 keV from the partner states discussed in the previous paragraph. Its angular momentum expectation value, $\langle \diamond | \hat{\jmath}_{n,z} | \diamond \rangle \simeq 0.5$, indicates that this level is dominated by the $K = 1/2^+$ substate of the $2d_{3/2}$ shell, despite it being directly connected to a $K = 5/2^+$ substate of the $1g_{7/2}$ shell on the prolate side of the Nilsson diagram. The reason for this hidden change in quantum number is the avoided level crossing slightly above $\gamma \simeq 20^\circ$. Coupling this neutron level to the proton level results in a $2^+$ or $3^+$ for the favored partner of the GM doublet, and an excited $3^+$ or $4^+$ for the excited state. 

These candidates for low-lying 2qp levels corroborate the assignment of quantum numbers to the low-lying levels made in Sec.~\ref{Sec:results}: the ground state assigned to $(6^-)$ and the isomer assigned to $(0^-)$ would be the GM partner levels obtained when coupling the  proton level from the $1g_{9/2}$ shell closest to the Fermi energy at the minimum of the false vacua to the negative parity neutron level just above the Fermi energy. The second isomer assigned to $(3^+)$ would be the lower of the two 2qp states that can be constructed when coupling the same proton level to the positive parity neutron level that is also located just above the Fermi energy.

We note that changing the spin assignment of the band head of the negative-parity band by one unit of angular momentum to $(6^-)$ also resolves a known anomaly for $^{114}$Rh that concerns the signature splitting in the odd-parity yrast band \cite{Liu11a,Navin2017}. Assuming it is built on a $6^-$, this band would then exhibit the same normal staggering as found for the negative-parity bands in lighter odd-odd Rh isotopes.

Although models today cannot accurately predict spins and parities when states differ by as little as 100 keV, our calculations nevertheless call into question the spin assignments of other neutron-rich odd-odd Rh isotopes. Triaxial ground states are predicted for all of these nuclei \cite{Hukkanen2023} and, thus, the relevant part of the Nilsson diagram presented in this work is expected to be qualitatively the same. As the proton shell is quite isolated in Fig.~\ref{fig:Nilsson}, one can expect that the differences in the structure of the low-lying states are due to the unpaired neutron.  

Focusing first on states with positive parity: for $^{104-112}$Rh, the expectation value of $\langle \hat{\jmath}_{n,z} \rangle$ for all the neutron shells of interest does not exceed $1.2$. When coupled with the isolated proton level, the GM doublet with largest possible angular momentum difference has spins $1^+$ and $5^+$, leaving little room to construct states with angular momentum 6 or above for these nuclei. This poses no problem for the $1^+$ and $5^+$ assignments for the ground and isomeric states in $^{104,108}$Rh~\cite{NUBASE20} but contradicts the $(6^+)$ assignments for the isomers in $^{106,110,112}$Rh. The latter are based on $\beta$-decay studies \cite{DeAisenberg1966,Takahashi1971,Lhersonneau1999}, but in all these cases the experimentally-observed $\beta$ feedings do not exclude $(5^+)$ assignments. A $(5^+)$ assignment in $^{110}$Rh might even explain a non-negligible feeding to the 1900-keV $(4^+)$ state in $^{110}$Pd \cite{Gurdal2012}. Considering that the long-lived high-spin states are anchor points for the spin-parity assignments based on measured transition multipolarities \cite{Porquet2002,Luo2004,Joshi2004}, our newly attributed spin-parity values induce changes to all other levels.

For the range of odd-odd isotopes between $^{104}$Rh and $^{112}$Rh, positive parity is assigned to the ground and isomeric states. The lowest high-spin negative-parity states are tentatively assigned as $(6^-)$ for $^{106-110}$Rh and $(7^-)$ for the heavier species \cite{ENSDF}. Such assignment is well-established for $^{104}$Rh by a magnetic moment measurement \cite{BizzetiSona1990,Vaman2004}, which points to a $\pi g_{9/2} \otimes \nu h_{11/2}$ configuration. All odd-odd Rh isotopes between $^{104}$Rh and $^{118}$Rh also exhibit a rotational band built on this negative-parity state. The similarity of the observed rotational bands was then typically used to justify the $\pi g_{9/2} \otimes \nu h_{11/2}$ interpretation for more neutron-rich isotopes, without specifying which of the deformed substates are involved \cite{Porquet2002,Fotiades03a,Porquet03a,Liu11a,Liu13a}. However, the low-lying structures are experimentally difficult to measure because of the presence of low-energy $\gamma$ rays. As pointed out earlier in Ref.~\cite{Fotiades03a}, this means that some transitions might have been missed. In addition, from $^{110}$Rh onward all the spins and parities are assigned based on systematics only. These observations suggest that the current assignments for both negative and positive parity states made in the databases \cite{NUBASE20,ENSDF} should not be relied upon and are urgent candidates for thorough future experimental evaluation. We also note that changing the systematic-based tentative spin assignment for the high-spin long-lived states in $^{116}$Rh and $^{118}$Rh would lead to normal splitting for the observed negative-parity band of these nuclei that so far are also assumed to be built on a $(7^-)$ band head \cite{Navin2017}. 

Based on our calculations and proposed spin-parity assignments, the existence of additional, but so far unobserved, low-lying isomers in the odd-odd $^{104-112}$Rh isotopes is unlikely. For $^{116}$Rh, however, the relevant neutron shells are the same as for $^{114}$Rh and a similar low-lying spectrum can be expected: one GM doublet of negative parity where the high-spin state is the ground state and one positive parity isomer having a spin in between those of the GM doublet. A previous experiment found only two long-lived states in $^{116}$Rh \cite{Hukkanen2023} but a doublet splitting below 20 keV would not have been resolved. While scarce, the available decay data also does not rule out a third long-lived $\beta$-decaying state \cite{Wang2001}.

\section{Summary and outlook}

In this Letter we have reported on the discovery of an additional isomeric state in $^{114}$Rh via Penning trap mass measurement using the PI-ICR technique. The ground state, as well as both the previously known and newly discovered isomers, were observed to be produced in proton-induced fission of $^{nat}$U while the $\beta$ decay of $^{114}$Ru only populates the isomers. Based on these results and other experimental data, we propose the $(6^-)$ ground state, the $(0^-)$ first isomer at $94.5(23)$~keV and the $(3^+)$ second isomer at $117.2(21)$~keV. Contrary to the literature \cite{NUBASE20}, the high-spin level is the ground state. We estimate the half-lives of both isomers to be about one second. These observations also explain the anomalous fission yield of $^{114}$Rh reported in Ref. \cite{Hukkanen2023}.

Although current modeling cannot provide reliable predictions for the order of states separated by as little as 100 keV, comparison to mean-field calculations with the BSkG3 model qualitatively support the proposed assignments by identifying the negative parity states as possible members of a Gallagher-Moszkowski doublet. As noted before in Ref.~\cite{Hukkanen2023}, accounting for the triaxial deformation of nuclei in this region is crucial to match the ground state of neighboring odd-mass Rh isotopes and the spectroscopic properties of nuclei in this region. Our work calls for significant additional experimental effort: first and foremost the decay schemes of $^{114}$Ru~\cite{Jokinen1992} and $^{114}$Rh~\cite{Lhersonneau2003} should be reinvestigated. Secondly, our theoretical analysis questions the assumptions behind several spin assignments in the level schemes of odd-odd $^{106-112}$Rh which should be revisited. Finally, BSkG3 hints at the possibility for a previously unobserved isomeric state in $^{116}$Rh.

During the review process we were made aware of an independent mass measurement of the long-lived states in $^{114}$Rh performed using the Canadian Penning Trap at the Argonne National Laboratory \cite{Liu2024}. The results reported in both works are consistent with each other.

\section*{Acknowledgments}

The authors would like to thank Dr. Marjut Hukkanen for inspiring this work and invaluable discussions.

The present research benefited from computational resources made available on the Tier-1 supercomputer of the F\'ed\'eration Wallonie-Bruxelles, infrastructure funded by the Walloon Region under the grant agreement No. 1117545. This project has received funding from the European Union’s Horizon 2020 research and innovation programme under Grant Agreements No. 771036 (ERC CoG MAIDEN) and No. 861198–LISA–H2020-MSCA-ITN-2019, from the European Union’s Horizon Europe Research and Innovation Programme under Grant Agreement No. 101057511 (EURO-LABS) and from the Research Council of Finland projects No. 295207, 306980, 327629, 354589 and 354968. W. R. is a Research Associate of the F.R.S.-FNRS and a member of BLU-ULB (Brussels laboratory of the Universe). J.R. acknowledges financial support from the Vilho, Yrj\"o and Kalle V\"ais\"al\"a Foundation. 

\bibliographystyle{elsarticle-num} 
\bibliography{bibfile}

\end{document}